\documentclass[iop]{emulateapj}
\usepackage{natbib}
\usepackage{pdfpages}

\newcommand{\msol}{\mbox{\,$M_\odot$}}        
\newcommand{\mum}{$\mu$m}

\begin{document}

\title{Starbursts are preferentially interacting: confirmation from the nearest galaxies}

\author{
Johan H. Knapen\altaffilmark{1,2}
\and Mauricio Cisternas\altaffilmark{1,2}
}
\altaffiltext{1}{Instituto de Astrof\'{i}sica de Canarias, V\'{i}a L\'{a}ctea S/N, E-38205 La Laguna, Spain}
\altaffiltext{2}{Departamento de Astrof\'{i}sica, Universidad de La Laguna, E-38206 La Laguna, Spain}

\begin{abstract}

We complement a recent ApJ Letter by Luo et al. by comparing the fraction of starburst galaxies which are interacting with the overall fraction of interacting galaxies in the nearby galaxy population (within 40\,Mpc). We confirm that in starburst galaxies the fraction of interacting galaxies is enhanced, by a factor of around 2, but crucially we do so by studying a sample of almost 1500 of the nearest galaxies, including many dwarfs and irregulars. We discuss how adjusting the starburst definition influences the final result and conclude that our result is stable. We find significantly lower fractions of interacting galaxies than Luo et al. did from their larger but more distant sample of galaxies, and argue that the difference is most likely due to various biases in the sample selection, with a representative sample of the nearest galaxies, such as the one used here, being the best possible representation of a general picture. Our overall conclusion is that interactions can and do increase the number of starburst galaxies, and that this result is extremely robust. By far most starburst galaxies, however, show no evidence of a present interaction.

\end{abstract}


\section{Introduction}

In current cosmological galaxy evolution models, interactions between galaxies play a particularly prominent role in shaping both the dark halos and the baryonic structure of modern-day galaxies. Almost equally important is enhanced star formation in the form of what is commonly called starbursts, which not only transforms large amounts of gas into stars in short timescales, but also produces substantial changes in the interstellar and intergalactic media, through winds and violent late stages of stellar evolution.

Starbursts are related to interactions in the sense that the latter can stimulate the former, and many starbursts are observed to occur in interacting or merging galaxies. There is abundant anecdotal and statistical evidence to back up these general statements (see, e.g., Toomre \& Toomre 1972, Larson \& Tinsley 1978, or the review by Schweizer 2005), and in particular in the case of the most extreme starbursts, such as the luminous and ultra luminous infrared galaxies (LIRGs and ULIRGs), it has long been clear that there is a strong statistical connection with galaxy interactions and mergers (e.g., Joseph \& Wright 1985, review by Sanders \& Mirabel 1996). Such extreme starbursts are very rare though, especially at the current cosmological epoch---there are no ULIRGs within some 70\,Mpc of us, for instance.

It is therefore vitally important to separate the anecdotal from the statistical when referring to the interconnections between starbursts and interactions. A large body of observational and numerical work has led to the general conclusions that indeed there is a causal connection, but also that statistically the increase in star formation rate (SFR) as a result of a galaxy-galaxy interaction or merger is limited (e.g., Larson \& Tinsley 1978; Bergvall et al. 2003; Kapferer et al. 2005; Di Matteo et al. 2007, 2008; Ellison et al. 2008, 2010, 2013; Robaina et al. 2009; Knapen \& James 2009; Rodrighiero et al. 2011; Saintonge et al. 2012; Moreno et al. 2015). In addition, it is clear that most interacting galaxies at present do not have an enhanced SFR at all (e.g., Knapen, Cisternas \& Querejeta 2015, hereinafter Paper~II). This may well be due to timescales, in the sense that such `quiescent' interacting galaxies may have had an enhanced SFR in the past or will have it in the future, but it is important to establish observationally, and confirm by simulations, the duration and intensity of a typical and possible `starburst' phase.

In this Letter, we report on the results from a study of the statistical connections between starbursts and interactions in a representative sample of 1500 of the most local galaxies, focusing on the fraction of starburst galaxies that is interacting, compared to the fraction of interacting galaxies in the overall galaxy population. This work forms an extension, and confirmation, of similar studies performed on more distant galaxies, often selected using the Sloan Digital Sky Survey (SDSS; e.g., Luo et al. 2014). It is a vital step towards understanding to what extent general conclusions on relations between starbursts and interactions, often reached on the basis of or influenced by studies of rare but striking objects, are applicable to the general galaxy population, consisting of significant numbers of dwarfs and other low-mass galaxies.

\section{Observational data, Sample, and Methodology}

We use a sample of 1478 nearby galaxies ($D<40$\,Mpc) selected from the {\it Spitzer} Survey of Stellar Structure in Galaxies (S$^4$G; Sheth et al. 2010). We use a number of key parameters, namely the SFR (from a combination of IRAS 60 and 100\,\mum\ fluxes; from Querejeta et al. 2015) and the SSFR (SFR divided by stellar mass, the latter from the dust emission-corrected {\it Spitzer} 3.6\,\mum\ images from the S$^4$G; also from Querejeta et al. 2015), and whether the galaxies are interacting (classes A---merging, B---highly distorted, and C---with minor distortions; from Knapen et al. 2014, hereinafter Paper~I).

As explained in more detail in Paper~II, for each sample galaxy we calculate the enhancement in its SFR and SSFR, $E({\rm SFR})$ and $E({\rm SSFR})$, by dividing these parameters by the median values for a control sample. The control sample is created for each galaxy individually, and consists of all those galaxies which are not interacting (and which also do not have a close companion---a further category defined in Paper~II but which we do not use in the current paper\footnote{The 138 galaxies in our sample that we classified as having a close companion in Paper~I (but which are not interacting) have the same behavior in terms of SFR and SSFR as the overall sample, which is why we limited our analysis to the galaxies in our interaction classes A, B and C.}), but which are close in morphological type (within $\pm1$ numerical class) and stellar mass (within $\pm0.2$ in $\log(M/M_\odot)$). 

Having calculated $E({\rm SFR})$ and $E({\rm SSFR})$, we define starburst galaxies as those which have values of $E$ above a certain number, in particular $E({\rm SFR})>5$ and $E({\rm SSFR})>4$. We show in the next Sections that these choices are reasonable, but also investigate the effects of using higher or lower limits, and of limiting the sample to certain ranges of stellar mass. 

To study the connection between starbursts and mergers we follow the same approach as Luo et al. (2014), namely determining how many of our galaxies defined as starburst are interacting (so in our categories A, B, or C), and then how many of all sample galaxies are interacting. The ratio between these two numbers, $B$, the starburst {\it boost} due to the interactions, is the main parameter reported in this Letter. For our full sample, we find that 18\% of our starburst and 9\% of all galaxies are interacting, leading to a boost factor $B=2.0$. We make further tests by reducing the sample of galaxies studied by imposing that they have a certain minimum stellar mass, as described below, but they do not alter our conclusions.

\section{Results}

\begin{figure*}
\centering
\includegraphics[width=5.5in]{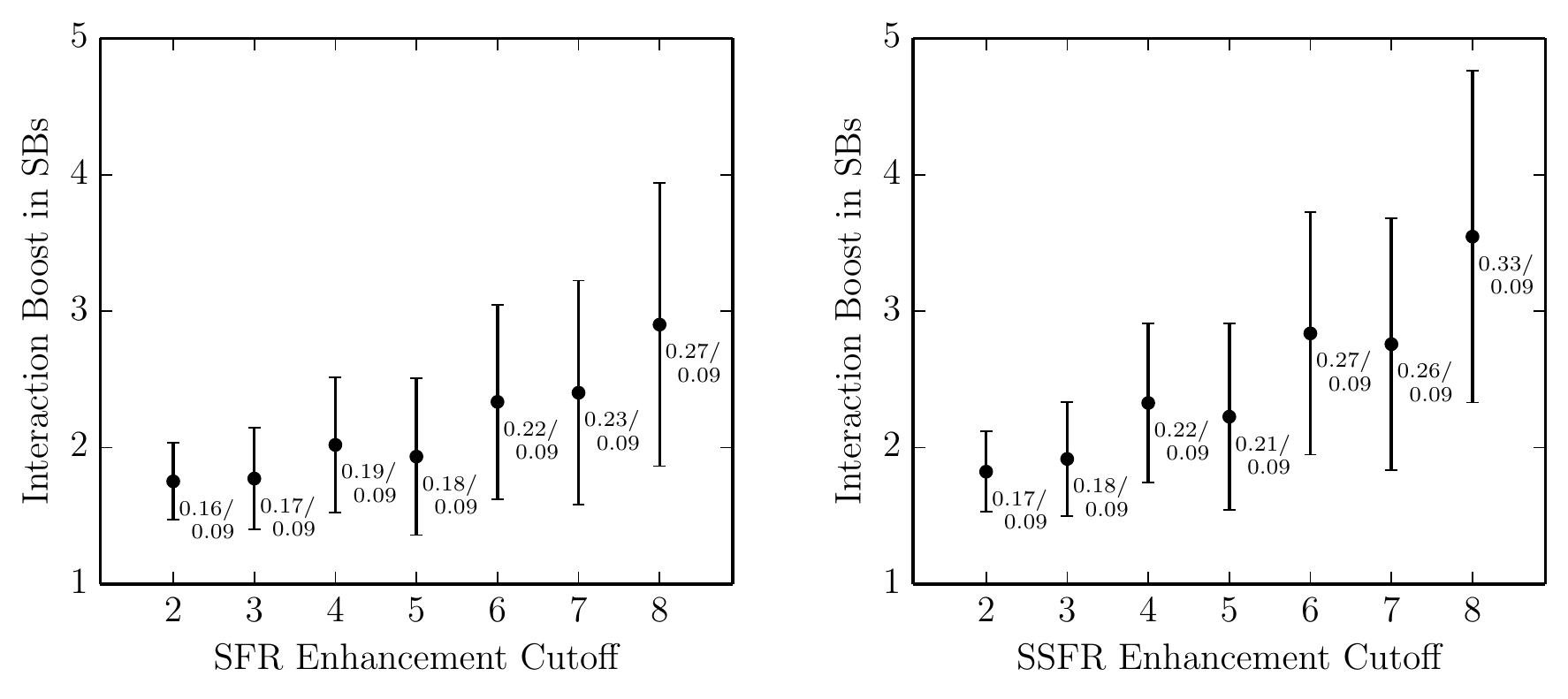}
\caption{Boost factor $B$ by which the fraction of interacting galaxies is enhanced among starburst galaxies as compared to the non-starburst control population, as a function of the enhancement cutoff used to define whether a galaxy is called a starburst: the factor $E$ by which the SFR and SSFR of a starburst is higher than those of their control sample. Small labels near each data point indicate the fraction of starburst galaxies that occur in interacting galaxies (top) and the fraction of all sub-sample galaxies that are interacting (bottom), the plotted data points indicate the ratio between these two numbers. The main conclusions of this Letter are based on reasonable values of $E({\rm SFR})=5$ and $E({\rm SSFR})=4$, which yield $B=2.0$ and 2.4, respectively.}
\label{SFR}
\end{figure*}

The parameter we are mainly interested in here is the enhancement of the fraction of interacting galaxies among the starburst population, as compared to the overall population of galaxies. For this, it is vital to use a reliable and meaningful definition of what a starburst is, and in Paper~II we argued that a reasonable definition is an enhancement of the SFR of a galaxy by a factor of 5 or more compared to its control population (a factor of 4 in SSFR). Using this value for $E$(SFR)$=5$, 18\% of starburst galaxies and 9\% of all galaxies are interacting. This implies a boost factor $B=2.0$ which is the main result reported here (for $E$(SSFR)$=4$, the numbers are 22\%, 9\%, and $B=2.4$). 

To investigate how robust these results are with changing starburst definitions, we show in Fig.~\ref{SFR} how $B$ varies with different values of our starburst criteria: the enhancement cutoffs $E$ for SFR (left panel) and SSFR (right panel). As might be expected, we see that the difference in interaction fraction between starbursts and the complete sample ($B$) increases as the starbursts become more extreme (higher $E$). But what Fig.~\ref{SFR} also shows very clearly is that the boost factor $B$ is robust, and in fact does not vary by more than some 25\% ($\pm0.4$ and $\pm0.6$ in SFR and SSFR, respectively) over the whole range of starburst definitions probed. And this range is rather large: $E=2$ hardly discriminates starbursts at all from star-forming galaxies, whereas $E=8$ selects only the most extreme starbursts (for comparison, Table\,1 in Paper~II lists the only 18 galaxies in our sample of 1478 with $E\geq10$). The conclusion from this test is that our result is robust, and does not depend significantly on the exact value used to define a starburst, nor on whether we use SFR or SSFR for the starburst definition.\footnote{We note here, as we do in Knapen \& James 2009 and in Paper~I in more detail, that other definitions of starbursts can lead to significantly different populations of galaxies. We use here the definition of SFR and SSFR enhancement, as do Luo et al. 2014, rather than alternatives such as, for instance, gas depletion time, absolute values of SFR or SSFR, or such absolute values normalized by area.}

\section{Discussion}

\subsection{Starbursts are preferentially associated with interactions---Yet most starbursts don't interact}

The main conclusion we reach is that the fraction of starburst galaxies that are interacting is significantly higher than in the whole population. The difference is a factor of 2, and this enhancement is stable even when changing the starburst definition (see Fig.\,\ref{SFR}). This number is very similar to that reported in the recent Letter by Luo et al. (2014), even though the properties of their input sample and the fractions of starburst and interaction galaxies among their sample are very different (see next Subsection). 

Our study of a sizeable sample of 1478 of the most local galaxies of all types thus confirms that indeed the fraction of galaxies associated with galaxy-galaxy interactions and mergers is significantly higher among starburst galaxies than among the general population of galaxies. And this in turn is further evidence that indeed mergers and interactions are among the main causes for the occurrence of starbursts in galaxies. We intentionally write here ``among the main causes'', as in spite of the factor of two difference in interaction fraction between the starburst and overall samples, the fraction of starburst galaxies, 22\% when using a reasonable definition of what constitutes a starburst, is still modest. The enhancement in this fraction related to interactions and mergers is undeniable, but most starburst galaxies, by far, still occur in galaxies without any evidence for interactions. This may well be related to timescales, with both the starburst phase and the visible stage of morphological evidence for interactions being relatively short-lived, but does mean that caution must be applied when discussing starbursts, interactions, and their interrelations. 

\subsection{Starburst fractions depend on sample biases}

\begin{figure*}
\centering
\includegraphics[width=5.5in]{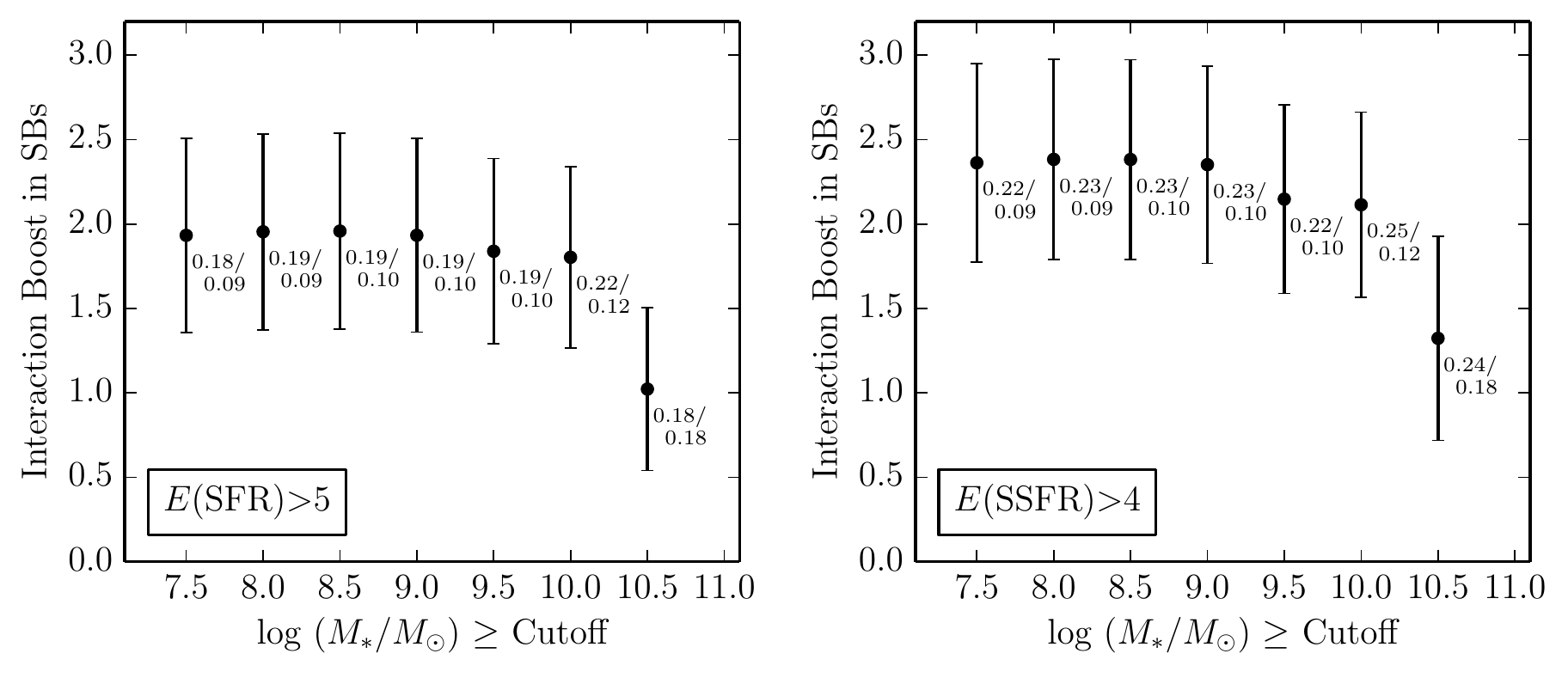}
\caption{As Fig.~\ref{SFR}, but now as a function of the stellar mass cutoff (lower limit) used to define the input sample.   The $B$ values in Fig.\,\ref{SFR} are reproduced as the leftmost points. All for $E({\rm SFR})=5$ (left panel) and $E({\rm SSFR})=4$ (right panel).}
\label{Mass}
\end{figure*}

Although our ``boost factor'', of $B=2$, by which starbursts are preferentially interacting compared to the control population is remarkably similar to that found by Luo et al. (2014), one very obvious difference between their results and ours is that their fractions of interacting galaxies are significantly higher. Luo et al. report that around 50\% of their starburst galaxies show evident merger features, against just below 20\% for their control sample. In contrast, we find values of 22\% and 9\%, respectively. 

We suspect that this difference is due primarily to biases introduced by the sample selection. There are several important differences in this respect, including

\begin{itemize} 

\item Luo et al. (2014) select galaxies at redshifts from 0.01 to 0.20, with an apparent magnitude cutoff of $r=17.72$, whereas our galaxies are within $D=40$\,Mpc and with $m_B<15.5$. This translates into limits in absolute magnitude of roughly $-17.5$ for our sample, and a range from $-17.8$ to $-23.2$ for Luo et al.

\item The starburst definition is similar in both papers, with SFRs that are $\sim5$ times larger than the median SFR of a control sample. But in practice, given the ``main sequence'' (Fig.~2 in Paper~II, for instance, or Fig.~1 in Luo et al. 2014) the SFR increases with stellar mass, and thus with absolute magnitude, so that starbursts will have higher absolute SFRs with higher masses. The starbursts, as well as the control galaxies, as defined by Luo et al. will thus have much higher SFRs.

\item In both cases, the presence of interactions is partly based on visual inspection of optical images. This, however, will give substantially different results in our case of very nearby galaxies which are well resolved in the imaging used, and the much more distant galaxies in the sample of Luo et al. (2014).

\end{itemize}

Many of these biases are hard to quantify, and it is almost impossible to correct the interaction fractions for them. But one test that we can easily perform is to simulate a different range in stellar mass (and thus in absolute magnitude and, in an indirect way, distance) in our sample. We show the results of this in Fig.~\ref{Mass}, which indicates how both the boost factor $B$ by which the interaction fraction increases in starburst galaxies (data points) and the interaction fractions among the starburst and control samples (small numbers near data points) change substantially when changing the mass range allowed in the sample. We see that the interaction fractions increase at higher-mass samples, but, in apparent contradiction to the constancy of the boost factor of around 2 between the work of Luo et al. (2014) and ours, $B$ drops in the highest mass cutoff bin. Here we must caution though for small subsample sizes, as our highest-mass subsample ($M_*>10.5\,\msol$) contains only 292 galaxies (of which 22 are starbursts and 270 non-starbursts, 4 and 48 of which are interacting, respectively).

The main conclusion here is, in any case, that in spite of these differences in sample selection and methodology, and possibly resulting biases, the overall result is extremely robust: interactions boost the fraction of starbursts, and do this by a factor of around $2-2.5$. The fact that we reproduce this result from the largest sample of the most nearby galaxies now available is very significant, and lends important further support to the conclusion that indeed interactions can and do increase the number of starburst galaxies (although most starbursts are not presently interacting).  

\subsection{Starbursts and interactions in low-mass galaxies}

\begin{figure}
\centering
\includegraphics[width=3in]{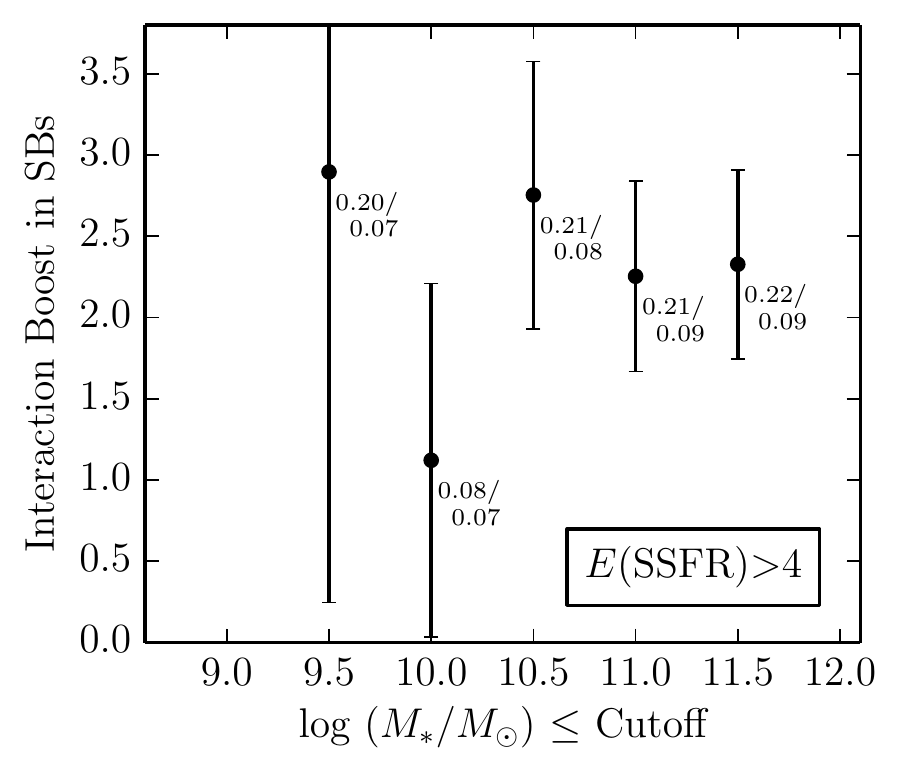}
\caption{As Fig.~\ref{Mass}, but now using an upper limit for the stellar mass as cutoff. This highlights the behaviour of low-mass galaxies (rather than the high-mass galaxies in Fig.~\ref{Mass}). Only the plot for $E({\rm SSFR})=4$ is shown because for the low-mass (e.g., dwarf) galaxies the SFR is a considerably less useful parameter than the SSFR. No significantly different behaviour among the lower-mass galaxies is seen.}
\label{Dwarfs}
\end{figure}

Fig.~\ref{Dwarfs} is like Fig.~\ref{Mass}, but using the stellar mass as an upper limit rather than the lower limit of Fig.~\ref{Mass}. There, we wanted to simulate the lack of low-mass galaxies included in samples at higher redshifts, but because our sample contains also low-mass galaxies, we can use a similar analysis to investigate whether the behavior of those galaxies with the lowest masses is statistically different from the main population.

Fig.~\ref{Dwarfs} shows that this is not the case: there is no significant different between the boost factor (by which starbursts are preferentially interacting compared to the control population) for low-mass galaxies, where the lowest stellar masses we can study here are $<10^{9.5}\msol$. Of course the uncertainties are large because only a few of these galaxies in our sample are defined as starbursts (only 5 of 367 of the $<10^{9.5}\msol$ galaxies, 1 and 25 of which are interacting, respectively), but the lack of evidence or a different behavior qualitatively agrees with what has recently been reported by, e.g., Lelli et al. (2014), Moreno et al. (2015) or Stierwalt et al. (2015). 

\section{Conclusions}

We present a detailed study quantifying to which extent starburst galaxies are preferentially interacting, using a sample of almost 1500 of the nearest galaxies for which we have reliable information on their SFRs, stellar masses, and on whether they are interacting with neighboring galaxies. A crucial difference between this study and others in the literature is that we base our results on a representative sample of local galaxies, including many of relatively low stellar mass. 

We confirm that among starburst galaxies the fraction of interacting galaxies is enhanced, by a factor of around 2, compared to the general population of local galaxies. Adjusting the starburst definition---the exact threshold by which the SFR or SSFR of a starburst galaxy must be enhanced compared to its control sample---allows us to conclude that our final result is stable. We find significantly lower fractions of interacting galaxies than Luo et al. (2014) did from a larger but more distant sample of galaxies selected from the SDSS survey, and argue that the difference is most likely due to various biases in the sample selection, with a representative sample of the nearest galaxies, such as the one used here, being the best possible representation of a general picture. Our overall conclusion is that interactions can and do increase the number of starburst galaxies, and that this result is extremely robust. By far most starburst galaxies, however, show no evidence of a present interaction.

We acknowledge financial support to the DAGAL network from the People Programme (Marie Curie Actions) of the European Union's Seventh Framework Programme FP7/2007-2013/ under REA grant agreement number PITN-GA-2011-289313, and from the Spanish MINECO under grant number AYA2013-41243-P. This research made use of the NASA/IPAC Extragalactic Database which is operated by JPL, Caltech, under contract with NASA.

\end{document}